\documentclass[useAMS,usenatbib]{mn2e}

\def\dg{$^{\circ}$}
\def\msun{\hbox{$M_{\odot}$}~}

\def\rsun{$R_{\odot}$}
\def\msun{$M_{\odot}$}

\usepackage{graphicx}  
 \usepackage{times}
  \usepackage{lscape}
  \usepackage{rotating}
  \usepackage{txfonts}

\title[The SMC B-type supergiant AzV322]{The SMC B-type supergiant AzV322: a g-mode pulsator with a circumstellar disc}
\author[Mennickent et al. ]
  {R.E. Mennickent$^{1}$\thanks{E-mail: rmennick@astroudec.cl},  
  Z. Ko{\l}aczkowski$^{2}$,
     I. Soszy{\'n}ski$^{3}$, M. Cabezas$^{1}$, 
       H.E. Garrido$^{4}$
\\
  $^1$Universidad de Concepci\'on, Departamento de Astronom\'{\i}a,
      Casilla 160-C, Concepci\'on, Chile\\
      $^{2}$  Instytut Astronomiczny Uniwersytetu Wroclawskiego, Kopernika 11, 51-622 Wroclaw, Poland  \\
       $^{3}$ Warsaw, University Observatory, Al. Ujazdowskie 4, 00-478 Warszawa, Poland \\ 
    $^{4}$ Instituto de Astronomia, Geof\'isica  e Ci{\^e}ncias Atmosf\'ericas, Universidade de S{\~a}o Paulo, Rua do Mat{\~a}o 1226, Cidade Universit\'aria, 05508-900 S{\~a}o Paulo, Brazil\\
       }
\date{}

\begin{document}


\maketitle 

\begin{abstract} 

We present a photometric and spectroscopic study of AzV322, an emission line object located in the Small Magellanic Cloud previously classified between O9 and B0.
We analyze 17.5 years of $I$ and $V$ band OGLE-II, III and IV light curves and find four significant frequencies, viz.\, 
$f_1$= 0.386549 $\pm$ 0.000003, $f_2$= 0.101177 $\pm$ 0.000005, $f_3$= 0.487726  $\pm$ 0.000015 and $f_4$= 0.874302 $\pm$ 0.000020 c/d.  
The $f_1$ frequency (period 2\fd58700 $\pm$ 0\fd00002) provides the stronger periodogram peak and gives a single wave light curve of full amplitude 0.066 mag in the $I$-band.  High-resolution optical spectroscopy confirms the early B-type spectral type and reveals prominent double peak Balmer, Paschen, OI\,8446 and He\,I\,5875 emissions. 
The spectral energy distribution shows significant 
 color excess towards long wavelengths possibly attributed to free-free emission in a disk-like envelope. 
Our analysis yields  $T_{eff}$ = 23\,000 $\pm$ 1500 K, log\, g = 3.0 $\pm$ 0.5, $M$ = 16 $\pm$ 1 \msun, $R$ = 31.0 $\pm$ 1.1 \rsun,  and  $L_{bol}$ = 10$^{4.87 \pm 0.06}$ $L_{\odot}$. 
AzV322 might be a member of the new class of slowly pulsating B supergiants introduced by \citet{2006ApJ...650.1111S} and documented by \citet{2007A&A...463.1093L}, however its circumstellar disk make it an hitherto unique object. 
Furthermore, we notice that a  O-C analysis for $f_1$ reveals quasi-cyclic changes for the times of maximum in a time scale of 20 years which might indicate a light-travel time effect in a very wide orbit binary with an undetected stellar component.

\end{abstract}


\section{Introduction}

AzV322 (SMC-SC9-161213, SMC113.86165, SMC725.06.36, SMC 2-6, LHA 115-S 36, LIN 406,  2MASS J01025999-7225395, $\alpha_{2000}$ = 01:02:59.99, $\delta_{2000}$ =  -72:25:39.1, B0e, $V$ = 13.73 mag, $B-V$ =  0.00 mag)\footnote{http://simbad.u-strasbg.fr/simbad/}
is a bright member of the Small Magellanic Cloud 
classified for the first time as H$\alpha$ emission object 
by \citet{1956ApJS....2..315H}, classification confirmed
by \citet{1975A&AS...22..285A} and \citet{1993A&AS..102..451M}. The star was
classified B0 ($V$ = 13.82, $B-V$ = -0.10, $U-B$ = -0.95) based on optical spectra in a list of probable members of the Small Magellanic Cloud \citep{1975A&AS...22..285A}
and photometrically as B0\,Ib  ($V$ = 13.90:, $B-V$ = -0.11:, $U-B$ = -0.97:, $E(B-V)$ = 0.12) by \citet{1970A&A.....9...95D}. Later, a classification of 
O9\,II was given based on the analysis of two low-resolution IUE spectra \citep{1997AJ....114.1951S}. The star is included in the group of massive short-period SMC stars with extra variability by  \citet{2014A&A...562A.125K}. A photometric period of 2\fd586981  is given by these authors based on the analysis of 8-years of OGLE-III data. A finding chart for the star and its surrounding stellar field is shown in Fig.\,1.

During our search for interacting binaries with Be-type components in the OGLE-II database, 
AzV322 was suspected to be a Double Periodic Variable (DPV), i.e. a semidetached 
Algol-like binary with a long photometric cycle lasting roughly 33 times the orbital period  \citep[eg.][]{2003A&A...399L..47M, 2016MNRAS.455.1728M,2017SerAJ.194....1M}.
This, and the fact that the star was the brightest of the SMC  DPV candidates, motivated the spectroscopic monitoring reported in this paper. However, the DPV classification
turned to be spurious since the long period was limited to the time baseline of the OGLE-II database, but not present when considering new photometric time series with longer time baselines. In addition, the first spectra we got showed well developed double emission lines which are unusual for an early B-type supergiant but typical of a classical Be star. All these results pointed to a complex picture  which is explored further in the present work. 

Let's remember that classical Be stars are rapidly rotating non-supergiants B-type stars that show or have shown Balmer line emission in the past \citep{2013A&ARv..21...69R}. The emission is formed in a circumstellar disk by ionization and subsequent recombination of the circumstellar material, mostly neutral hydrogen. The origin of the Be star phenomenon has remained elusive for more than a century,  but recent research based on high signal to noise photometry performed with satellites suggests that it is linked to the presence in the stellar surface, of gravity-mode oscillations producing difference frequencies with larger amplitudes than the original ones \citep{2016arXiv161101113B}. According to these authors, {\it " ... significant dissipation of pulsational energy in the atmosphere may be a cause of mass ejections from Be stars"}. These non-periodic mass ejections, documented in many multi-wavelength studies, are the origin of the circumstellar disk-like envelope characterizing classical Be stars. In spite of these promising advances, more investigation is needed to clarify how Be stars eject mass into the circumstellar medium. 

On the other hand, recent evidence for gravity-mode oscillations driven by opacity was also reported in a sample of periodically variable B-type {\it supergiants}, where the cause
for non-vanishing g-mode pulsations, which were expected in principle to be damped, has been attributed to  its reflection in the hydrogen-burning shell, which significantly reduces radiative damping in the core
\citep{2007A&A...463.1093L, 2006ApJ...650.1111S}. On the contrary to Be stars, B-type supergiants do not show evidence for a circumstellar disk, but they show P-Cygni type H$\alpha$ emission revealing line formation in an expanding atmosphere, a radiatively driven wind. 

In this paper we report pulsations in the B-type supergiant AzV322, which in principle can be attributed to gravitationally-supported oscillation modes. We will show that, in spite of its supergiant classification, AzV322 has attributes of classical Be stars: double peak emission revealing a disk-like envelope and infrared excess compatible with free-free emission in the disk. 
The study of AzV322  might  be relevant in the context of pulsations in hot supergiants and the origin of the Be star phenomenon. 

The paper is organized as follow: in Section 2 we introduce the photometric datasets used in our analysis,  details of our spectroscopic observations are given in Section 3, in Section 4 we present our results including the light curve analysis, the study of the spectroscopic data and the spectral energy distribution,  in Section 5 a discussion is provided along with a possible interpretation for the system and finally  our conclusions are given in Section 6.

\section{Photometric data}

The photometric time-series analysed in this paper are taken from the OGLE project databases.
We include OGLE-II data \citep{2005AcA....55...43S}\footnote{http://ogledb.astrouw.edu.pl/$\sim$ogle/photdb/} and OGLE-III/IV data\footnote{ 
OGLE-III/IV data kindly provided by the OGLE team.}. The OGLE-IV project is described by \citet{2015AcA....65....1U}.
The whole dataset  consists of  1529 $I$-band magnitudes and 161 $V$-band magnitudes taken during a time interval of 17.5 years.
A summary and characterization of these datasets is given in Table\,1.

When studying the spectral energy distribution,  we included broad-band photometry extracted from different sources with the aid of the VizieR photometric tool \footnote{http://vizier.u-strasbg.fr/vizier/sed/doc/}. 
A compilation of these fluxes and magnitudes is given in Table\,2. In the table, when  multiple fluxes were present in a given filter, we list
the measurement with less error and closer to the documented stellar position.

\begin{figure}
\scalebox{1}[1]{\includegraphics[angle=0,width=8cm]{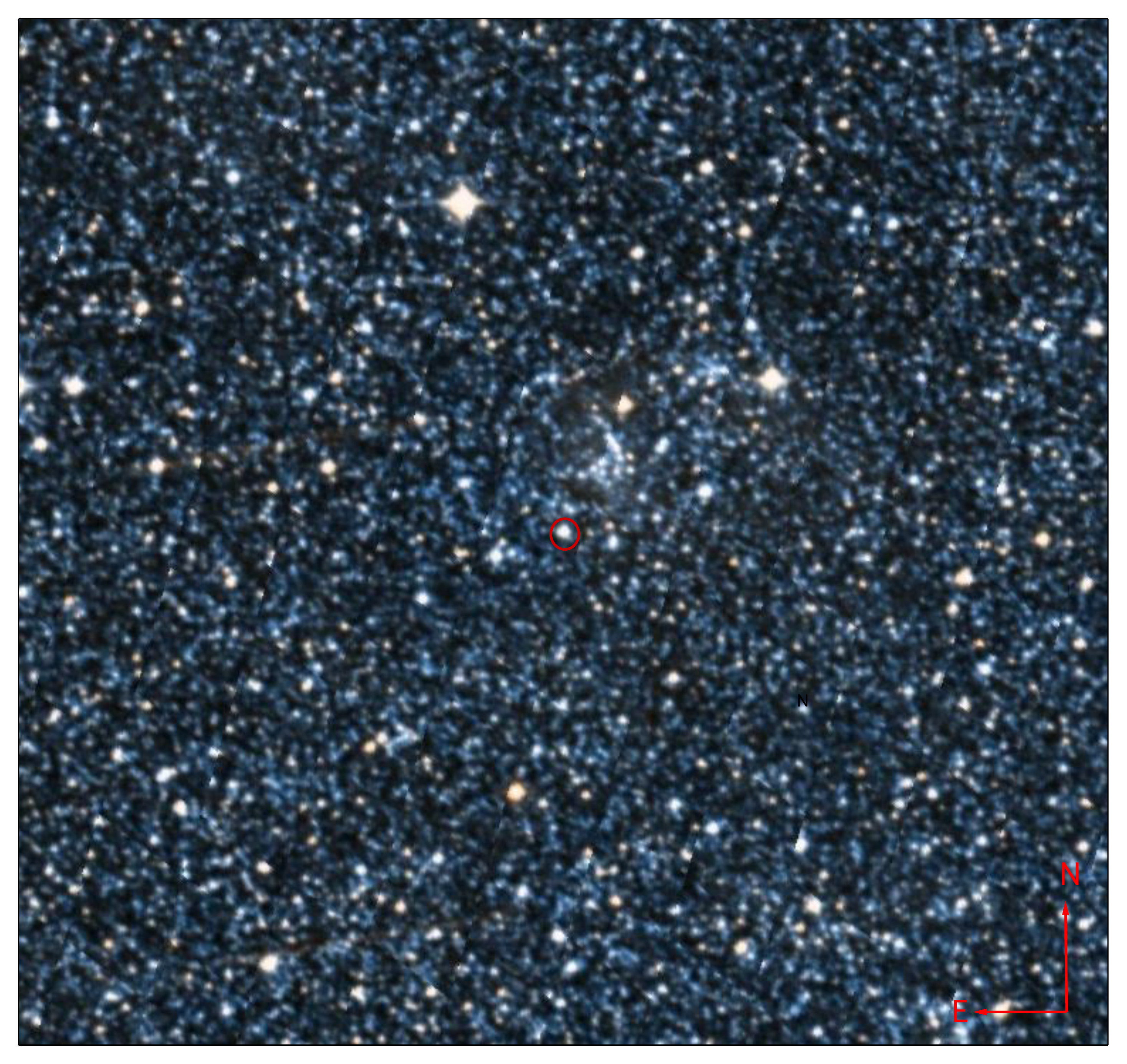}}
\caption{The stellar field around AzV322 spanning 
a field of view of 11.5 x 11.5  squared arc minutes; the image, from ESO-MAMA, has a resolution of 0.67 arcsec/pix. 
AzV322 is shown by a circle at the center of the image.  
Two open clusters are at the north-west location, OGLE\,233 and OGLE\,234.
}
  \label{x}
\end{figure}

\begin{table}
\centering
 \caption{Summary of survey photometric observations. The number of measurements, starting and ending times for the series and average magnitude and standard deviation (in magnitudes) are given.
 Single point uncertainties in the $I$-band and $V$-band are between 4 and 6 mmag.}
 \begin{tabular}{@{}lcrrccc@{}}
 \hline
Database &N &$HJD_{start}$ &$HJD_{end}$&mag &std. &band \\
\hline
OGLE-II  &326& 621.86357 &1873.69863&13.70 &0.03 &$I$\\
OGLE-III &      724& 2104.88015& 4953.92420&13.71 &0.04 &$I$\\
OGLE-IV  & 479& 5346.93180 &7312.65138 &13.77 &0.03 &$I$\\
OGLE-II &   33& 665.92289 &1510.59748 &13.82 &0.03 &$V$\\
OGLE-III  & 94& 3326.61986 &4954.92914 &13.81& 0.04 &$V$\\
OGLE-IV  & 34 &5399.87964 &6601.58580 &13.88 &0.04& $V$\\
 \hline
\end{tabular}
\end{table}

\begin{table}\label{Flux}
 \centering
\caption{Fluxes and their errors derived from magnitudes reported by the VizieR photometric tool.}
\begin{tabular}{ccccc}
\hline
Filter                  & $\lambda$    &    $f_\lambda$   &$\sigma$\,$f_\lambda$ \\
                          & ($\mu$m)              &    (Jy)             &        (Jy)         \\\hline    
Johnson $U$         &  3.53E-1    &    15.2E-3             &    0.4E-3            \\                        
Johnson  $B$        &  4.44E-1    &    13.7E-3             &    1.2E-3    \\
Johnson $V$         &        5.54E-1      &    11.7E-3             &    0.8E-3    \\
Johnson $J$        &        1.25E+0      &    5.57E-3             &    0.09E-3    \\
Johnson $H$       &        1.63E+0      &    3.96E-3             &    0.11E-3    \\
Johnson $K$        &        2.19E+0      &    3.45E-3             &    0.11E-3    \\
Cousins $I$           &        7.89E-1      &    9.09E-3             &    0.80E-3    \\
DENIS $I$             & 7.90E-1              &    8.91E-3        &    0.79E-3    \\
SDSS $I$              & 7.63E-1          &    7.78E-3           &    0.34E-3    \\
SDSS $r$             & 6.25E-1          &    9.27E-3           &    0.45E-3    \\
SDSS $g$             & 4.82E-1          &    11.3E-3           &    0.5E-3    \\
2MASS $J$        & 1.24E+0          &    5.33E-3           &    0.13E-3    \\                                    
2MASS $H$        & 1.65E+0         &    4.20E-3           &    0.11E-3    \\
2MASS $K_s$  & 2.16E+0          &    3.57E-3           &    0.11E-3    \\
WISE $W1$  & 3.35E+0          &    1.86E-3           &    0.04E-3    \\
WISE $W2$  & 4.60E+0          &    1.29E-3           &    0.03E-3    \\
WISE $W3$  & 1.16E+1          &    5.81E-3           &    79E-6    \\
WISE $W4$  & 2.21E+1          &    6.10E-3           &    0.73E-3    \\
IRAC 1   & 3.55E+0              &    1.79E-3        &    0.04E-3    \\
IRAC 2  & 4.49E+0          &    1.48E-3        &    0.02E-3    \\
IRAC 3   & 5.73E+0             &    1.10E-3         &    0.03E-3        \\
IRAC 4  & 7.87E+0           &    763.0E-6        &    37.0E-6        \\
\\\hline
\end{tabular}
\end{table}

\begin{figure*}
\scalebox{1}[1]{\includegraphics[angle=0,width=18cm]{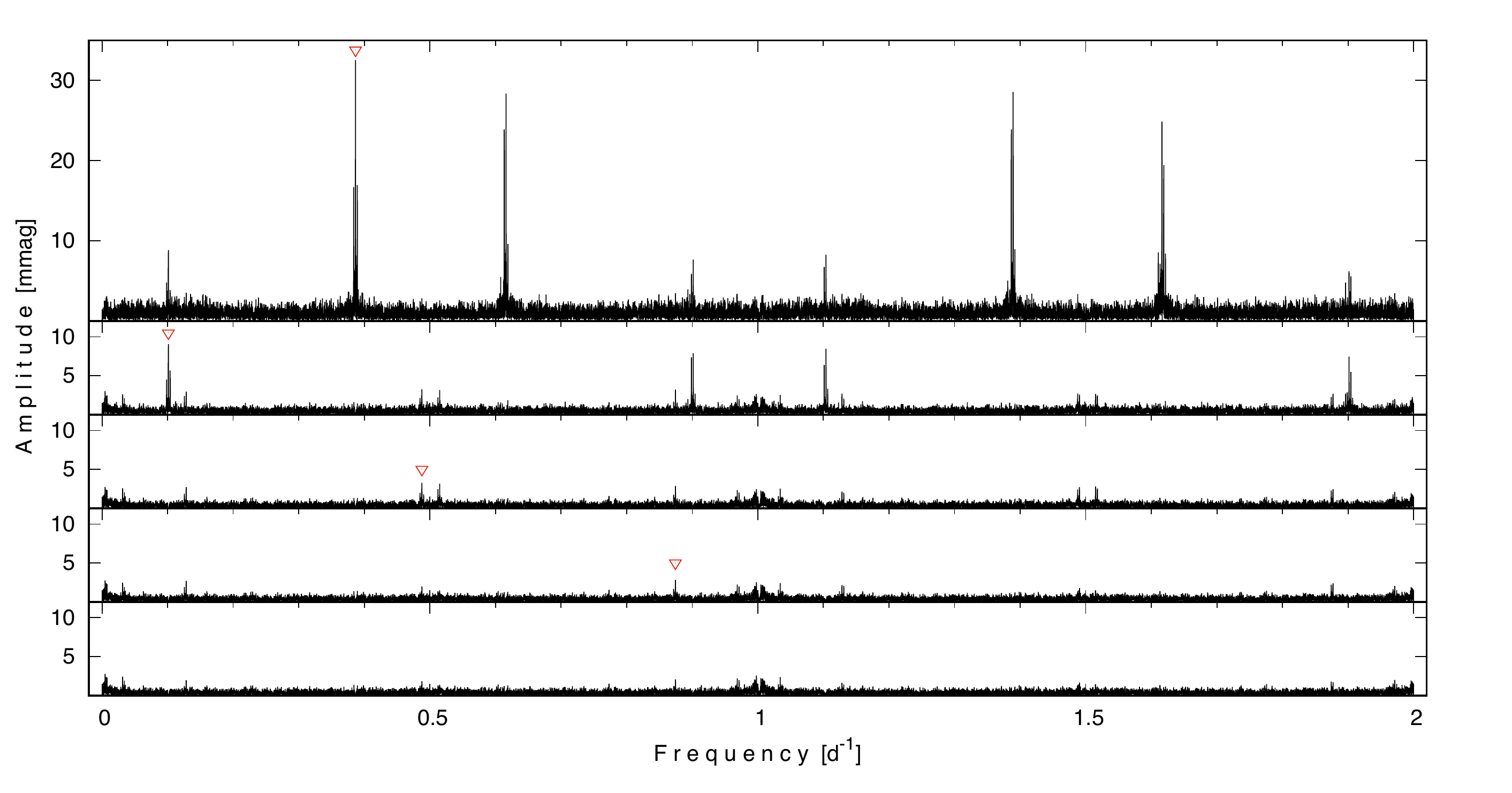}}
\caption{The periodograms showing the whitening process. From up to down the triangles show the frequencies $f_1$ to $f_4$.}
  \label{x}
\end{figure*}

\section{Spectroscopic data}

Spectra were taken between November 2006 and August 2009 with the 
echelle spectrograph mounted at the  Ir\'en\'ee du Pont  2.5m telescope\footnote{http://www.lco.cl/telescopes-information/magellan/telescopes-information/irenee-du-pont/instruments/}  and the {\it Magellan Inamori Kyocera Echelle}
(MIKE\footnote{http://www.lco.cl/telescopes-information/magellan/instruments/mike}) mounted in the 6.5m  Clay telescope 
in Las Campanas Observatory.  
The du Pont spectra have a wavelength range of 3940$-$7490 \AA~
and resolving power 40\,000. The MIKE double echelle spectrograph provided wavelength coverage of
3390$-$4965 \AA~(blue camera) and 4974$-$9407 \AA~ (red camera) 
with resolving power also of 40\,000. 
The spectra were reduced and calibrated with \texttt{IRAF} \citep{1993ASPC...52..173T}. The result of the this process was a set of continuum-normalized wavelength-calibrated, one-dimensional not flux-calibrated spectra. A summary of these observations is given in Table\,3.

\begin{table}
\centering
 \caption{Summary of spectroscopic observations.The heliocentric julian day (HJD - 245 0000) at mid-exposure  is given. 
 The $S/N$ ratio is calculated in the continuum around 4000 \AA. 
 $\Phi$  refers to the phase according to the 9\fd8837  period calculated according to the ephemerides given in Eq.\,1.}
 \begin{tabular}{@{}lccccc@{}}
 \hline
Instrument &UT-date &exptime (s) &$S/N$ &HJD &$\Phi$\\
\hline
MIKE &2006-09-26 & 	360	&35 &4004.70866	&0.672\\
MIKE & 2006-10-02& 	600	 &40 &4010.69670	&0.279 \\
MIKE &2006-11-26 	&300 &15	&4065.69103	&0.847\\
MIKE &2007-07-20 & 	1110&30	&4301.63093	&0.738\\
MIKE &2007-11-08 &	6300&45	&4412.52745	&0.967\\
MIKE &2007-11-09 & 	1000&30&	4413.51920&	0.068\\
MIKE &2008-01-05 & 	600	&40&4470.51848	&0.839\\
echelle &2009-08-25 &	2000&15 &	5068.80851&	0.422\\
echelle &2009-08-25 &	2000&15 &	5068.83272&	0.424\\
 \hline
\end{tabular}
\end{table}

\section{Results}

 \subsection{Analysis of the light curve}
 
We removed the long-term tendency in the $I$-band subtracting from the data a spline-function. 
Since the $V$-band light curve consisted of much less datapoints, it was corrected shifting the OGLE-II and OGLE-IV data to the average of the
OGLE-III dataset. We investigated the resulting $I$-band light curve performing a Fourier analysis with the {\rm Period04}\footnote{https://www.univie.ac.at/tops/Period04/} software \citep{2005CoAst.146...53L} between 0 and 2 c/d. A dominant frequency $f_1$= 0.386549 is found in the periodogram, and a subsequent whitening
process reveals three additional statistically significant periodic signals, viz.\,  $f_2$= 0.101177,
$f_3$= 0.487726 and
$f_4$= 0.874302
(Figs.\,2, 3 and 4).
After the process of subtraction of the four frequencies 
the residual light curve only shows a
bunch of very low frequencies and their
aliases around 1 c/d.
This is typical for objects with an irregular
(erratic) variability at the level of 0.0025 mag.
The results of this process are given in Table\,4. 

Considering the whole dataset, we find the following ephemerides for the maximum of the short-cycle:


  \begin{equation}
 HJD = 2454620.3747(55) + 2\fd587000(17)\,E 
  \end{equation}

Since the dominant frequency $f_1$  has
small side-peaks, likely indicating disturbances in 
the periodic process (changes of period, amplitude
or another parameters), we divided the light curve in seasons for further analysis, which is possible due to the fact that the OGLE project observes
the SMC in longer than half year seasons.
 The time series data with removed signal of $f_2$ were divided
to one-season time-series. For each season we did the fit
with fixed frequency $f_1$ (obtained from the non-linear fit
to whole data set), but for every season we estimated amplitude and time of maximum (phase). With these results we can
obtain the O-C diagram and see how much the amplitude
of $f_1$ is variable; it shows substantial changes over a long
time scale (Fig.\,5). Furthermore, the times of maximum change quasi-cyclically in a time scale of 20 years, a scale
comparable with the data time baseline (Fig.\,6). We notice that the maximum deviations in the O-C diagram are +0.08 and -0.1 cycles or
+0.21 and -0.26 days.    
We did the same analysis for $f_2$ and find that
the amplitude of $f_2$ is too small and the O-C diagram is too
noisy; the points are distributed inside 2$\sigma$  therefore we considered the variability not significant (Fig.\,7). 
Subsequent frequencies $f_3$ and $f_4$ produce light curves of lower amplitude hence no significant result can be obtained with the above analysis
for these frequencies.

To test if the above results were sensible to the method used for removing the long-term tendencies we performed 
several tests with different methods and parameters
of long-term variability subtraction, for instance we used spline
fits for each season separately (and with careful inspection on all steps). We confirm that the results are basically the same.
Sometimes we observe additional low-amplitude frequencies, but $f_1$, $f_2$, $f_3$ and $f_4$
are always present in the Fourier analysis (i.e. they are real),
and they have always similar (and variable for $f_1$ and $f_2$) amplitudes.
We conclude that our light-curve analysis is robust and the results reliable.

  \begin{table}
\centering
 \caption{Results of the Fourier analysis of the $I$-band light curve. Frequencies, amplitudes (average in the case of $f_1$) and their errors are given. }
 \begin{tabular}{@{}cccc@{}}
 \hline
Label&frequency & period &amplitude  \\
 &(d$^{-1}$)  &(d) & (mag)  \\
\hline
$f1$&	0.386549 $\pm$ 0.000003&2.5870 &   0.0331 $\pm$ 0.0006 \\
$f2$&	0.101177 $\pm$ 0.000005&9.8837	&  0.0095 $\pm$ 0.0006 \\	 
$f3$&	0.487726  $\pm$ 0.000015 &2.0503	&  0.0034  $\pm$  0.0006\\
$f4$&	0.874302 $\pm$ 0.000020&1.1438	&  0.0027  $\pm$ 0.0007\\
\hline
\end{tabular}
\end{table}



\begin{figure}
\scalebox{1}[1]{\includegraphics[angle=0,width=8cm]{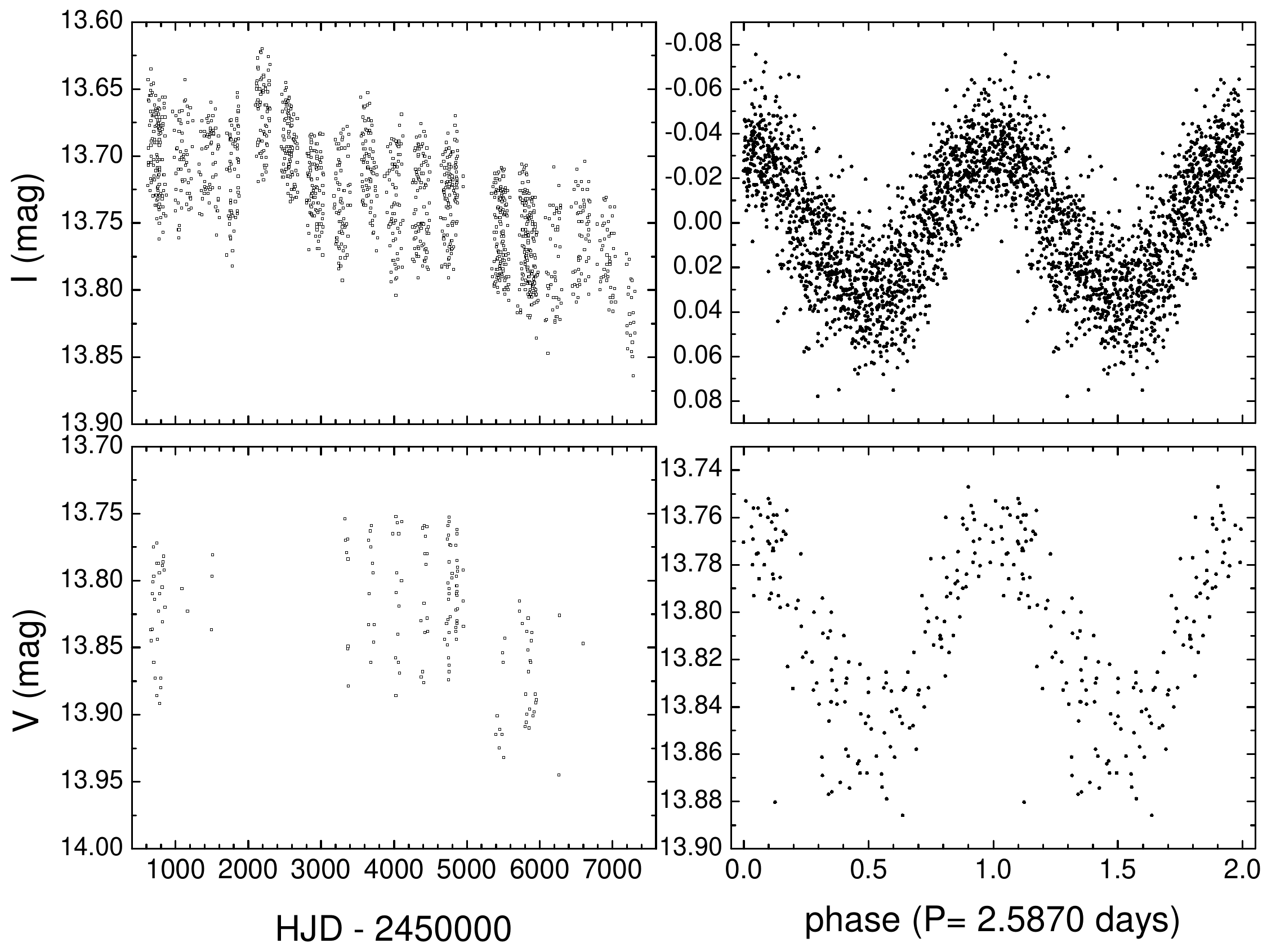}}
\caption{OGLE $I$-band and $V$-band light curves spanning 17.5 years (left) and the fit residuals phased with the short period of 2\fd5870 (right).
}
  \label{x}
\end{figure}

\begin{figure}
\scalebox{1}[1]{\includegraphics[trim= 1cm 1cm 1.5cm 1cm,angle=0,width=8cm]{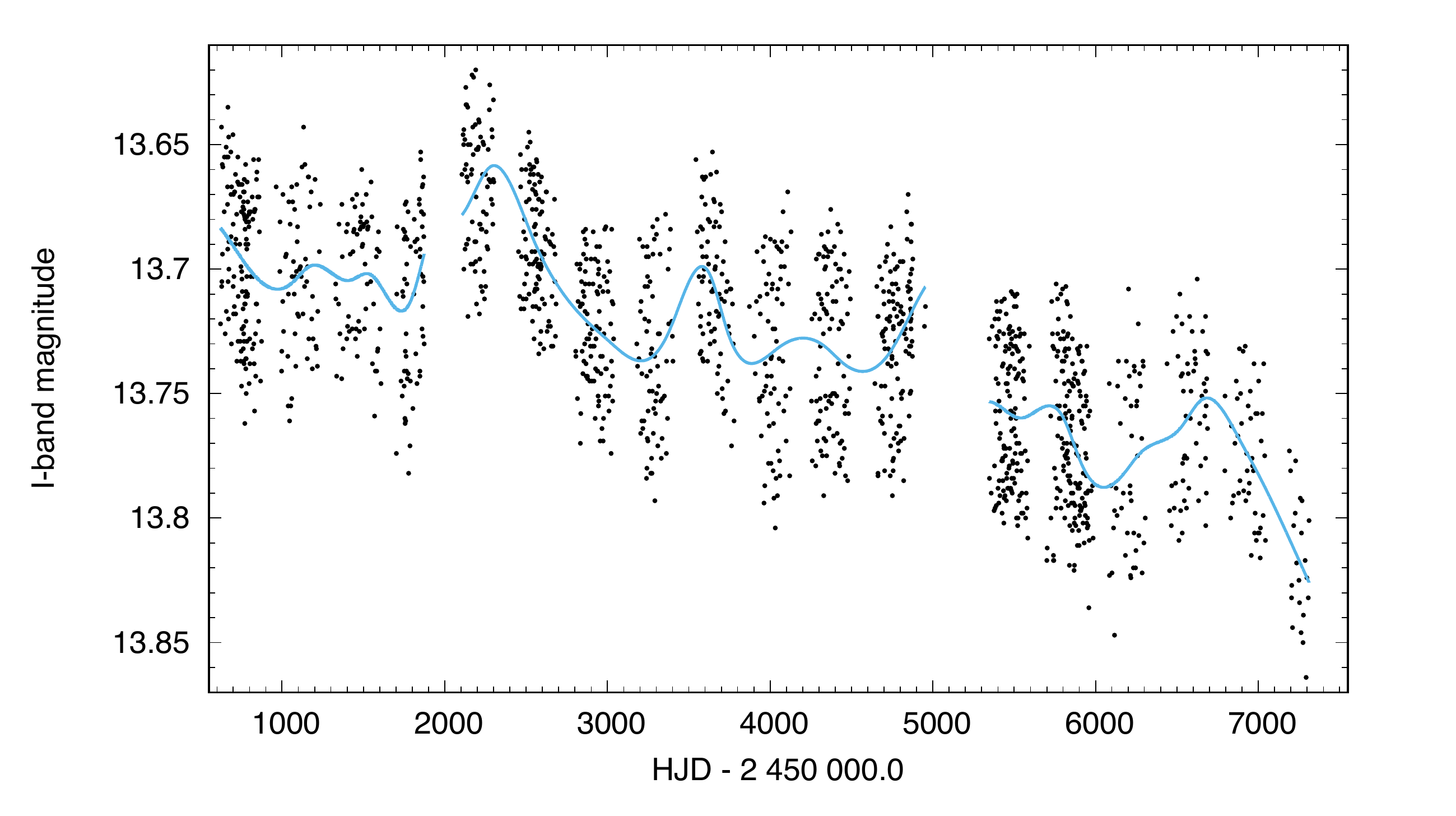}}
\caption{The OGLE $I$-band light curve and the spline-function used to remove the long-term tendency.
}
  \label{x}
\end{figure}

\begin{figure}
\scalebox{1}[1]{\includegraphics[trim= 2.5cm 1cm 2cm 3cm,angle=-90,width=8cm]{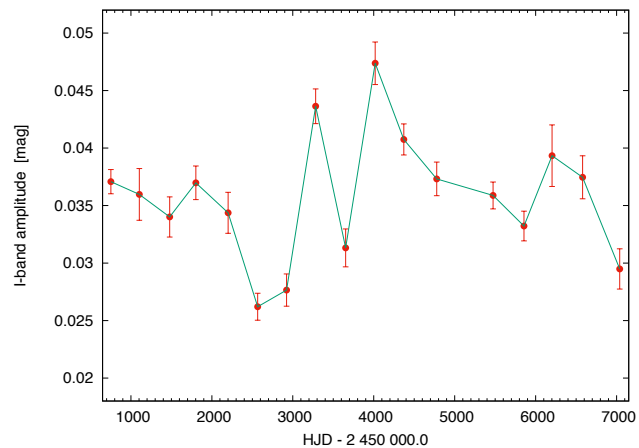}}
\caption{The seasonal amplitudes of the $f_1$ signal.}
  \label{x}
\end{figure}

In the rest of the paper we refer to $f_1$ as the short-cycle  and $f_2$ as the long-cycle frequency. We  notice that the amplitude of the short cycle is larger in the $V$-band (0.043 mag, estimated from the light curve) than in the $I$-band (0.033 mag, Table 4). 


\begin{figure}
\scalebox{1}[1]{\includegraphics[trim= 2.5cm 2cm 2.5cm 3cm, angle=0,width=8cm]{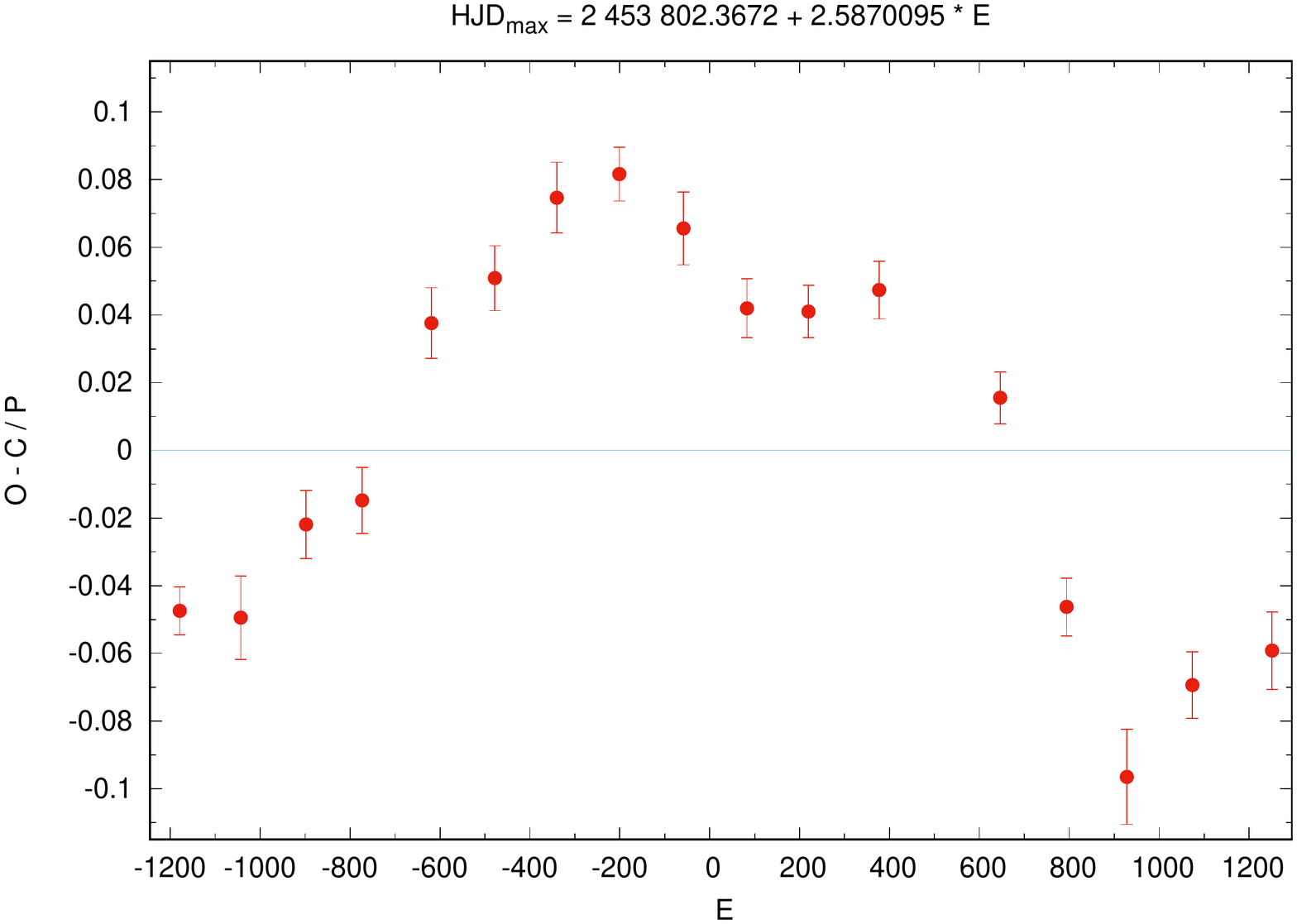}}
\caption{The observed minus calculated diagram for $f_1$.
}
  \label{x}
\end{figure}

\begin{figure}
\scalebox{1}[1]{\includegraphics[trim= 2.5cm 2cm 2.5cm 3cm, angle=0, width=8cm]{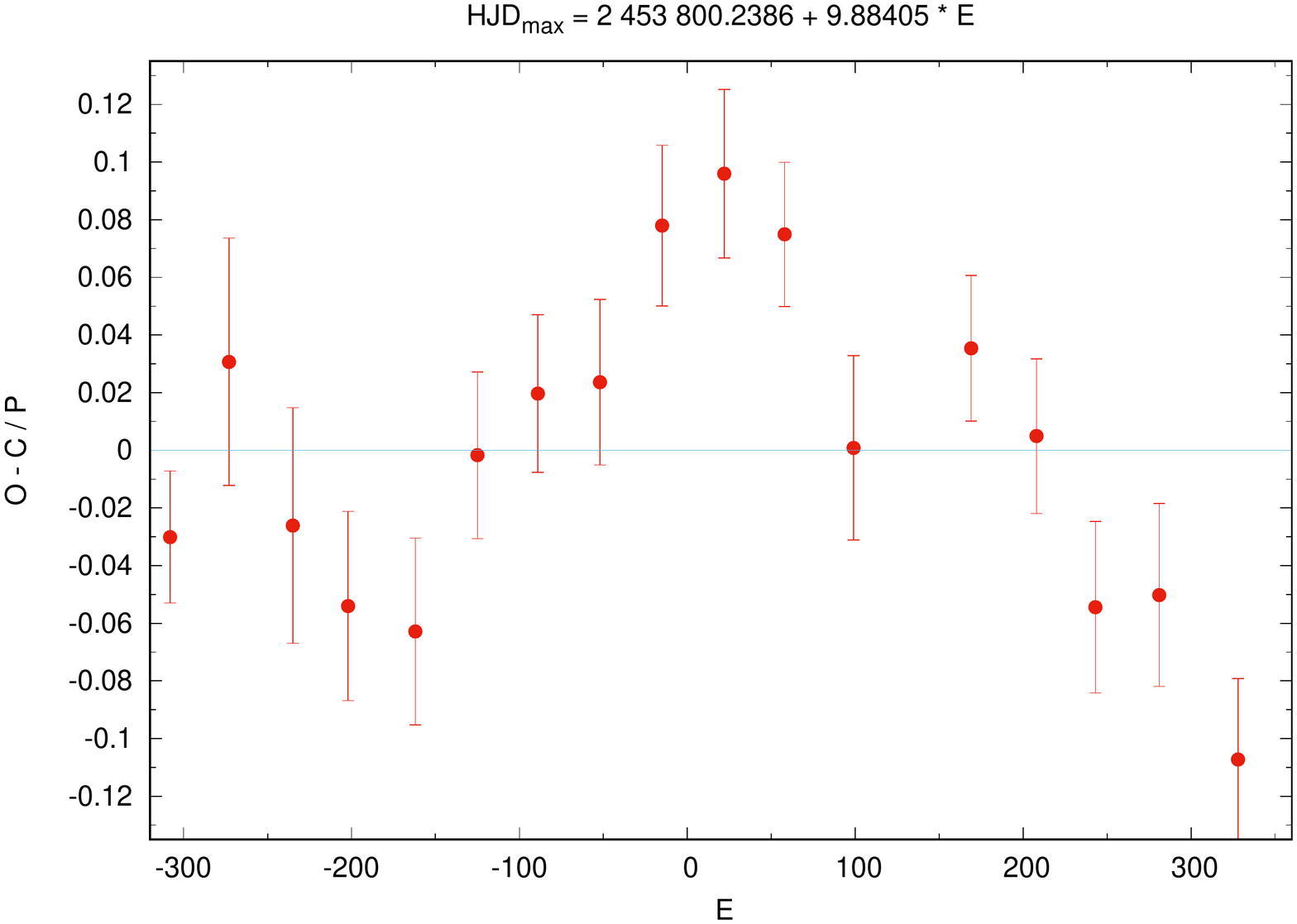}}
\caption{The observed minus calculated diagram for $f_2$.
}
  \label{x}
\end{figure}

\subsection{Analysis of spectroscopic data}

The O-type published in the literature is discarded since we do not find He\,II lines in our spectra. We don't observe either any forbidden emission line so the object is not a B[e] star.
 MgII\,4481 is very weak compared to HeI\,4471 indicating an early B-type star.
Looking in the line strength ratios He\,I 4009/4026 and He\,I 4121/4142 we estimate a spectral type B1, according to a comparison with the Gray spectral atlas\footnote{https://ned.ipac.caltech.edu/level5/Gray/frames.html}.

Average properties for the emission lines are presented in Table\,5, including the ratio between the violet and red peak intensity relative to the normalized continuum $V/R$ $\equiv$ $(I_V-1)/(I_R-1)$. We observe 
that the equivalent width of the H$\alpha$, H$\beta$ and H$\gamma$ emission lines remains almost constant and only small line profile variability is detected  (Fig.\,8).
The blue emission peak has practically the same intensity that the red emission peak and the peak separation increases with the Balmer series order. This gradient is  typical signature of a Keplerian Be star disk.  
We also observe in the near-infrared the Paschen series in emission, along with OI\,8446 (Fig.\,8). 
The mean velocity for the central absorption of Paschen lines shown in Fig.\,9  is 106 $\pm$ 9 km s$^{-1}$.  OI\,8446 
 is blended with HI\,8438 and this explains the large  $V/R$ = 1.50 $\pm$ 0.05.
We notice that the average spectrum reveals weak double emission in He\,I\,5875, with peak separation 246 $\pm$ 2 km s$^{-1}$ and central absorption with a minimum at 89.6 km s$^{-1}$. This emission is weak, with maximum intensity about 5\% over the continuum. 
Emission is not detected in any other helium line.
We measured for the  NaD\,1 line $EW$ = 0.160 $\pm$ 0.005 \AA\, suggesting $E(B-V)$ = 0.05 according to the calibration provided by Munari \& Zwitter (1997).

\begin{figure}
\scalebox{1}[1]{\includegraphics[angle=0,width=8cm]{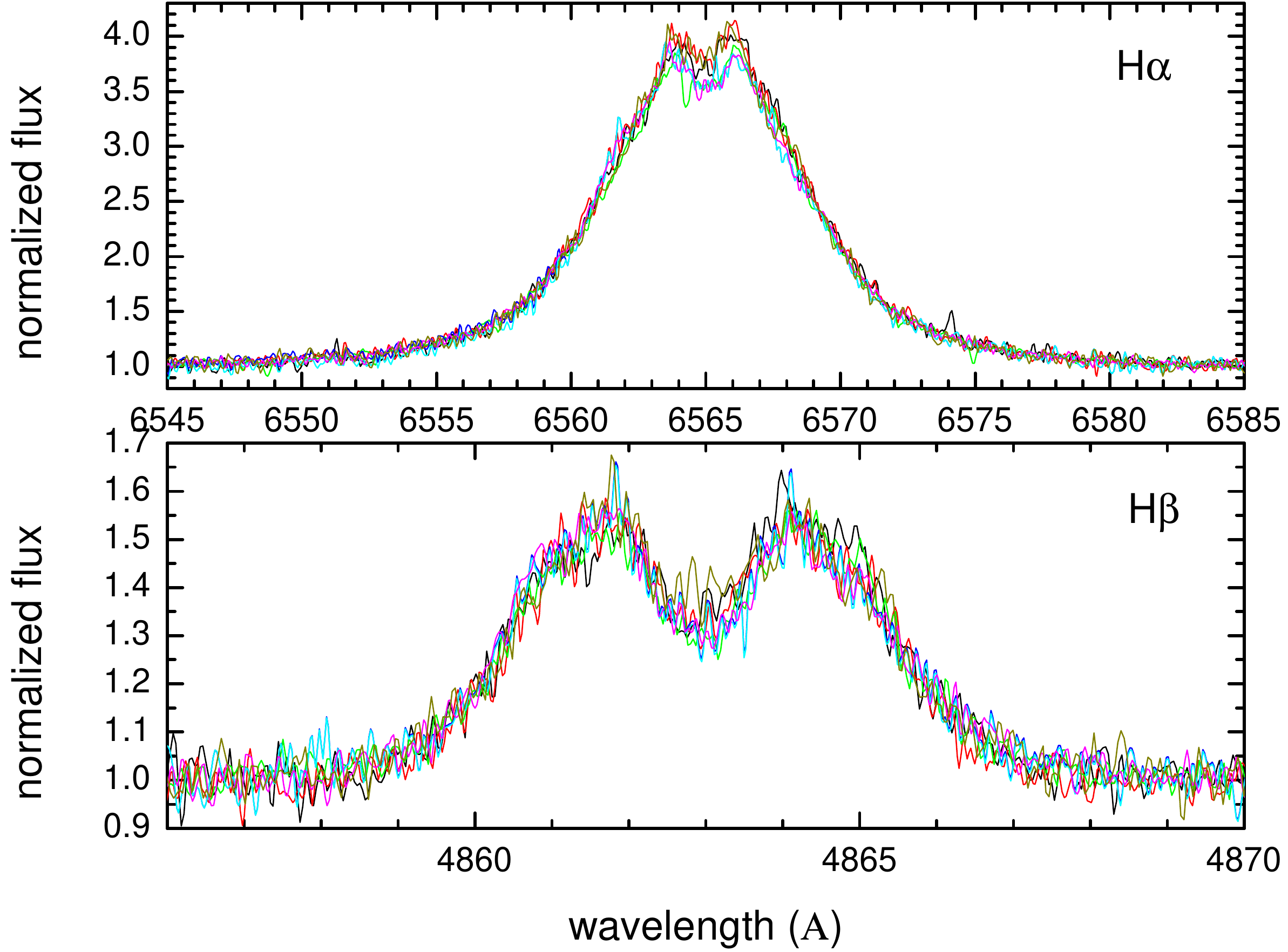}} \\
\scalebox{1}[1]{\includegraphics[angle=0,width=8cm]{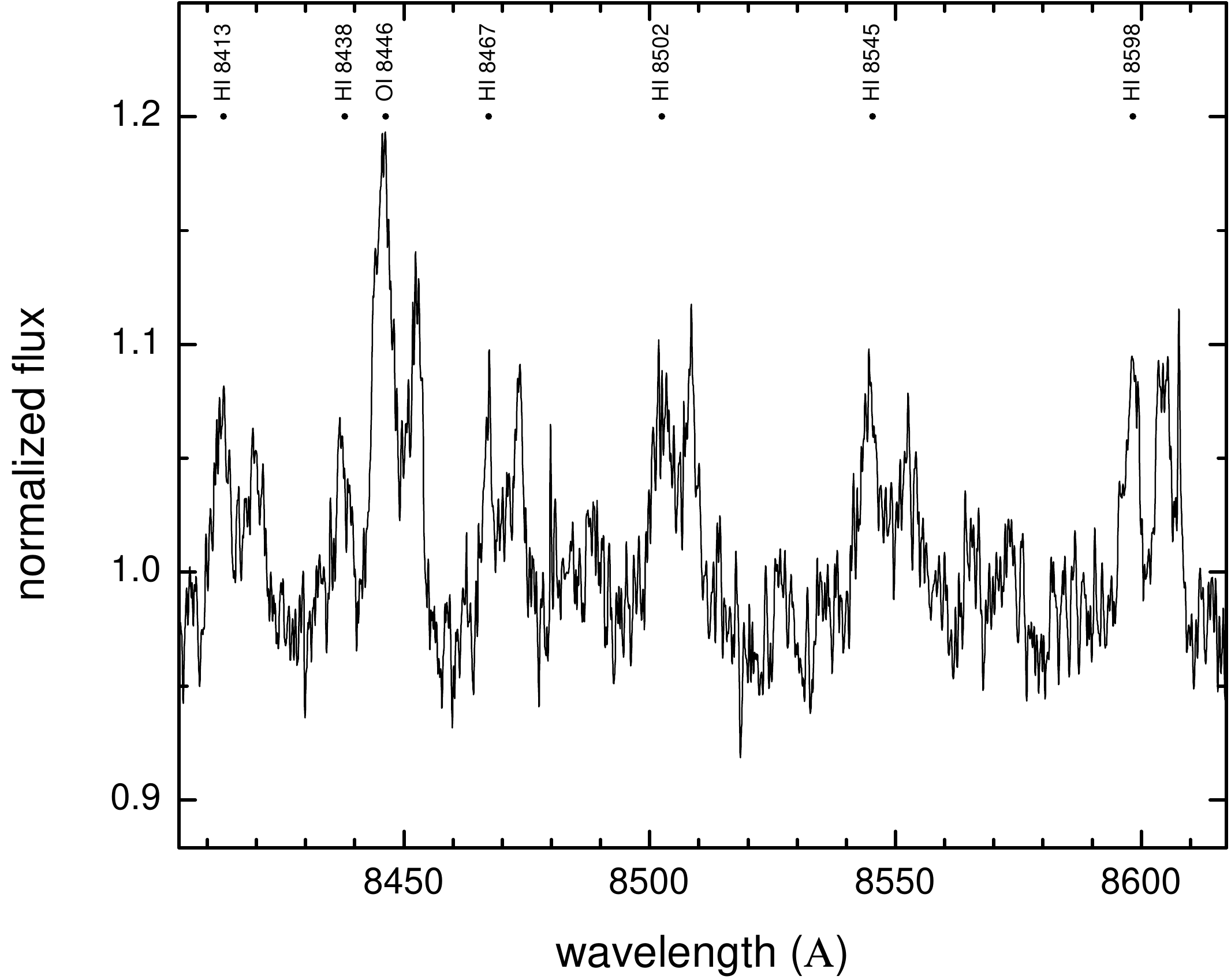}} \\
\scalebox{1}[1]{\includegraphics[angle=0, width=8cm]{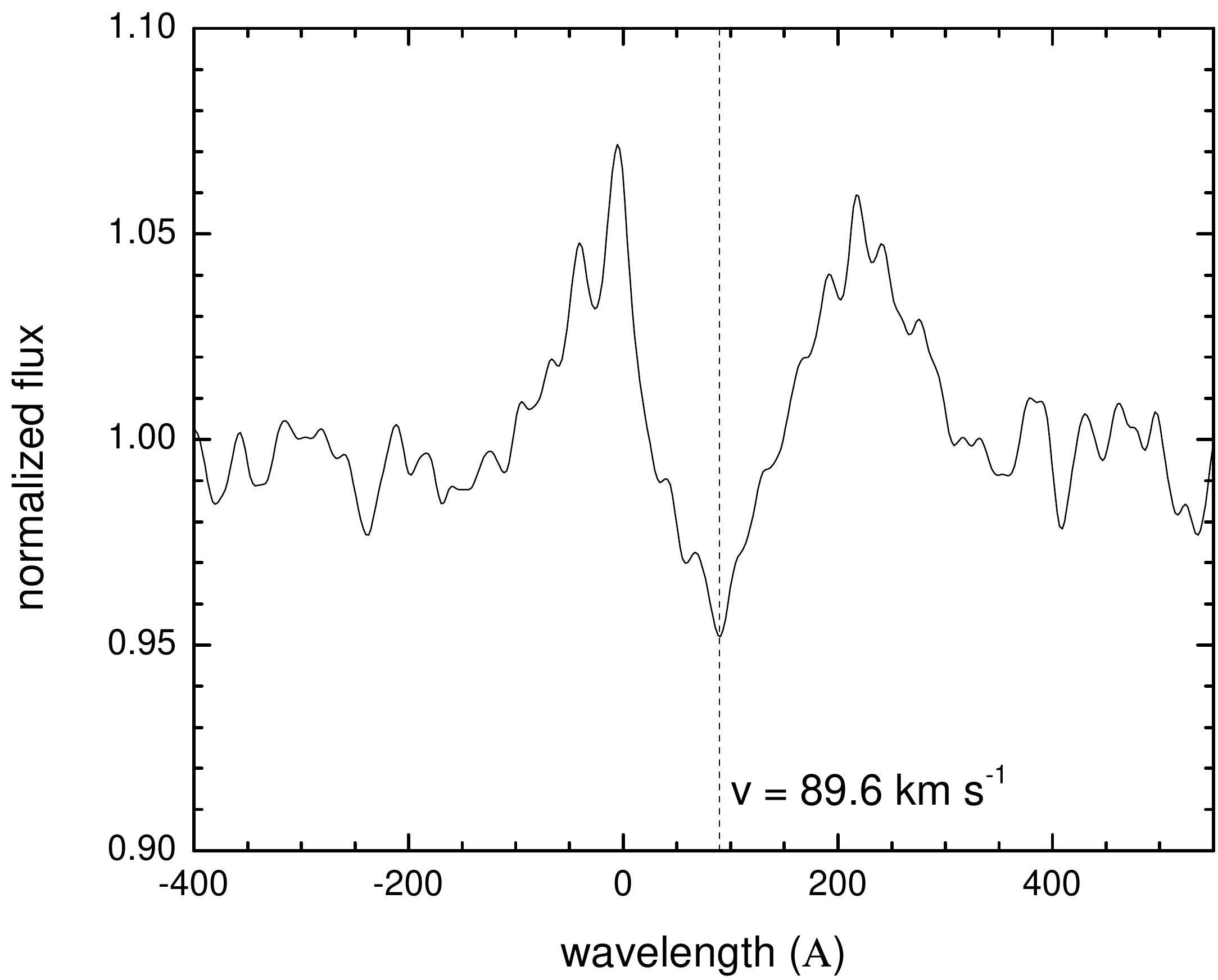}} 
\caption{(Up): H$\alpha$ and H$\beta$ profile variability for MIKE spectra. (Middle): 
A segment of the near-infrared spectral range in the MIKE mean spectrum. Hydrogen Paschen lines are labeled, along with OI\,8446.
(Down): The He\,I\,5875 line profile in the MIKE mean spectrum.  }
  \label{x}
\end{figure}

 We notice  than AzV322 might be a fast rotator, due to the broad absorption lines. However, we don't provide projected rotational velocity since 
 the low S/N of our spectra impeded to separate adequately the contributions of the 
macro-turbulent velocity from the projected rotational velocity, being the former important for early B-type supergiants  \citep{2006A&A...451..603D}.

The helium absorption lines in the single spectra are too noisy to provide reliable radial velocities for the Be-star photosphere.
In the MIKE averaged spectrum, the average radial velocity of the helium lines of Table\,6 is 96.5 $\pm$ 17.2 km s$^{-1}$ (std), just slightly lower than the mean velocity of the H$\alpha$ emission lines that is about 102 km s$^{-1}$. We attribute the large scatter in helium velocities to  (i): the quality of the data; noisy line profiles introduce artifacts  in the velocity determined with Gaussian fits and (ii) the presence of non-radial pulsations related to the photometric oscillations.  The situation is different for the emission lines, they have larger S/N ratio and their radial velocity errors are smaller, probably around 2 km s$^{-1}$, as derived from the scatter at a given epoch (Table\,7).


When phasing the emission line radial velocities with the short period no relation is observed, but  when phasing with the long period a smooth oscillation of amplitude about 18 km s$^{-1}$  is observed (Fig.\,9).
To construct this figure we have used the ephemerides for the long-cycle with the same zero point that for the short-cycle.
 


 \begin{figure}
\scalebox{1}[1]{\includegraphics[angle=0,width=8cm]{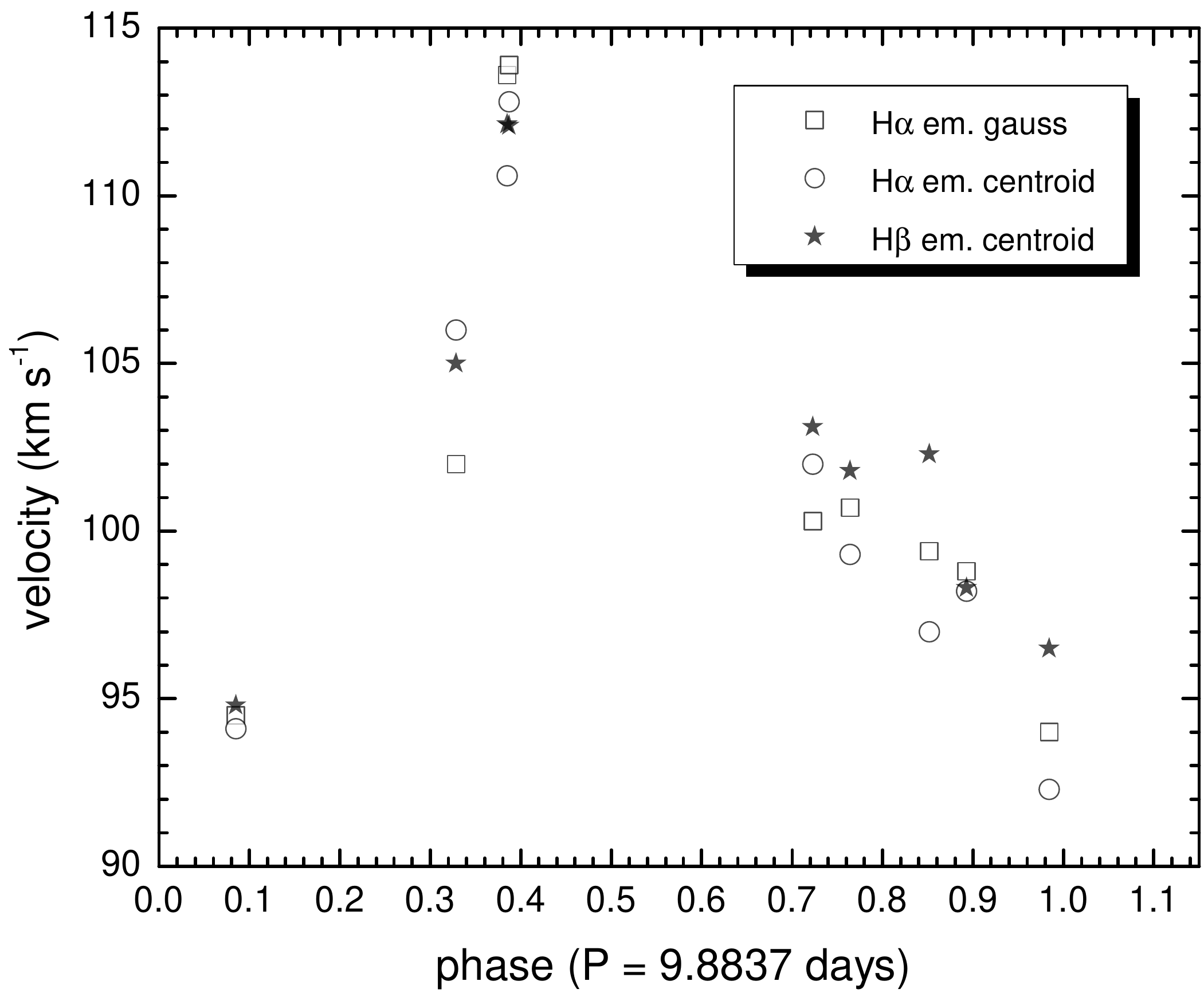}}
\caption{Radial velocities measured for different emission lines phased with the 9\fd8837 period.
}
 \label{x}
\end{figure}

\begin{table}
\centering
 \caption{Average measurements for emission lines and their standard deviations.
 Data for Paschen lines shown in Fig.\,8 are included, along with those for He\,I\,5875 of the same spectrum. The fifth columns gives
 the radius of the line forming region relative to those of He\,I\,5875 (see text for details). }
 \begin{tabular}{@{}lcccc@{}}
 \hline
Line &$EW$ &$\Delta \lambda$  &$V/R$& r \\
 & (\AA) & (km s$^{-1}$) & &\\
\hline
H$\alpha$ &-28.50 $\pm$ 1.10 & 95.1 $\pm$ 9.0 & 0.97 $\pm$ 0.05 &6.69\\
H$\beta$& -2.95 $\pm$ 0.46 & 159.3 $\pm$ 15.3 & 0.99 $\pm$ 0.03 &2.38\\
H$\gamma$ & -0.47 $\pm$ 0.08 &198.6 $\pm$ 5.1   & 0.98 $\pm$ 0.03&1.53 \\
Paschen &-0.80 $\pm$ 0.13 &227 $\pm$ 21 &  1.06 $\pm$ 0.27&1.17\\
He\,I\,5875 &-0.16 $\pm$ 0.02 & 246 $\pm$ 2 &1.15 $\pm$ 0.05 &1.00\\
\hline
\end{tabular}
\end{table}

\begin{table}
\centering
 \caption{  Radial velocity measured with a gaussian fit in the average spectrum.}
 \begin{tabular}{@{}lccc@{}}
 \hline
Line & wavelength (\AA) &RV (km s$^{-1}$) \\
\hline
He\,{\sc i} & 3819 & $83.0\pm1.0$\\
He\,{\sc i} & 4009 & $125.4\pm3.1$\\
He\,{\sc i} & 4026  & $92.9\pm1.0$\\
He\,{\sc i} & 4143 & $107.0\pm1.0$\\
He\,{\sc i} & 4387 &$92.1\pm1.0$ \\
He\,{\sc i} & 4922 &$78.5\pm 1.6$ \\ \hline
Mean  & - &$96.5\pm 17.2$\\
 \hline
\end{tabular}
\end{table}


\begin{table}
\centering
 \caption{Radial velocities of emission lines  measured with line gaussian fits or with the centroid of
 the emission line.  The typical error,  obtained from the scatter of velocities measured at a given phase, is 2  km s$^{-1}$. }
 \begin{tabular}{@{}cccc@{}}
 \hline
HJD& $RV_{\alpha}$ (km s$^{-1}$ )&$RV_{\alpha}$ (km s$^{-1}$ )&$RV_{\beta}$ (km s$^{-1}$ )  \\
&Gauss &Centroid  & Centroid  \\
\hline
2454010.69670&  102.0	&106.0 &  105.0 \\
2454065.69103 & 98.8& 	98.2   &98.3\\
2454301.63094 & 100.7&	99.3  &101.8\\
2454412.52745  &94.0	&92.3 & 96.5\\
2454413.51920 & 94.5	 &94.1  &94.8\\
2454470.51848  &99.4&	97.0 & 102.3\\
2454004.70866 & 100.3&	102.0 & 103.1\\
2455068.80851  &113.6&	110.6  &112.1\\
2455068.83272 & 113.9&	112.8  &112.1\\
\hline
Average $\pm$ std &101.9 $\pm$ 7.2 &101.4 $\pm$ 7.1 &102.9 $\pm$ 6.2 \\
\hline
\end{tabular}
\end{table}

 \begin{figure}
\scalebox{1}[1]{\includegraphics[angle=0, width=8cm]{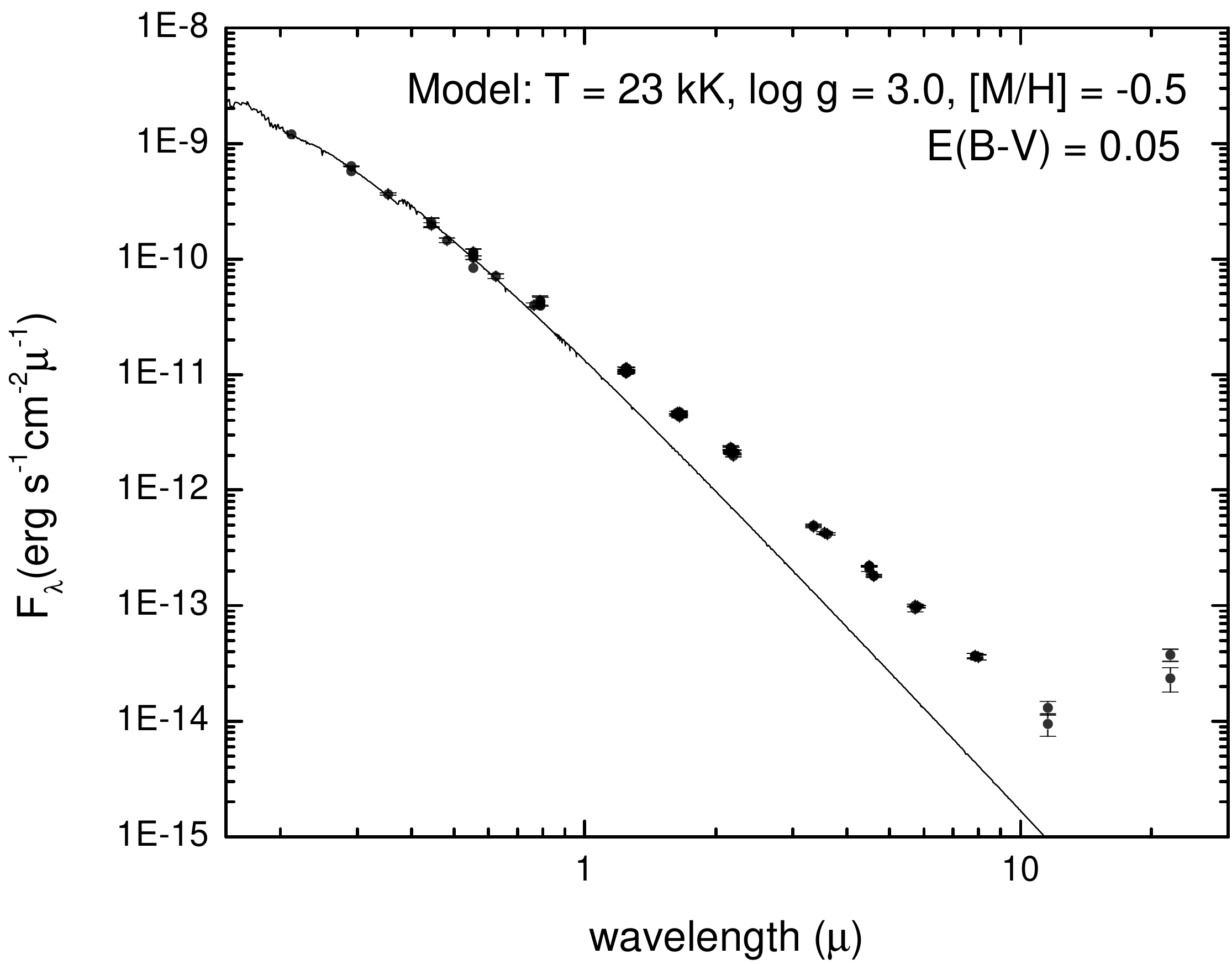}}
\caption{The spectral energy distribution and the best synthetic model.}
  \label{x}
\end{figure}

\subsection{Spectral energy distribution and stellar parameters}

The spectral energy distribution was constructed by compiling the published fluxes at different bands available in the VizieR SED viewer service\footnote{http://vizier.u-strasbg.fr/vizier/sed/}. A Marquant-Levenberg non-linear least square fit to the observed fluxes was done with a synthetic spectrum using the formula:\\

\begin{small}
$f_{\lambda , reddened}=  f_{\lambda} (R/d)^2  10^{-0.4E(B-V)[k(\lambda -V)+R(V)]}, $\hfill(1)\\
\end{small}
 
\noindent  
where  $f_{\lambda}$ is the stellar flux and $R$ the stellar radius,  $k(\lambda-V) \equiv E(\lambda-V)/E(B-V)$ is the normalized extinction
curve, $R(V) \equiv A(\lambda)/E(B-V)$ is the ratio of reddening to extinction at $V$ and $d$ is the distance to the system. In absence of any other better approach, we used the average Galactic Extinction Curve parametrized by Fitzpatrick \& Massa (2007) to calculate reddened fluxes. 
We used a synthetic spectrum characterized by T = 23 kK, log g = 3.0 and [M/H] = -0.5, from the grid of models given by Castelli \& Kurucz (http://wwwuser.oat.ts.astro.it/castelli/). Models with temperatures 18, 26 and 30 kK were also tested by failed to reproduce well the SED. We adopt as the system temperature T = 23\,000 $\pm$ 1500 K, that
 fits the effective temperature calibration for a B0.5 supergiant provided by \citet{2007A&A...463.1093L}.
The free parameters of the fit were $R/d$ and $E(B-V)$. 
The best fit, minimizing $\chi^{2}$,  was obtained considering fluxes with wavelengths shorter than 0.7$\mu$. It gave $(R/d)^2$= 1.33(4)E-22  and  $E(B-V)$= 0.046 $\pm$ 0.004  mag (Fig.\,10). This exercise confirms the color excess determined from the equivalent width of the NaD\,1 line and shows the evident color excess at wavelengths longer than 0.7$\mu$. It also gives the stellar radius; considering $d$ = 60.6 $\pm$ 1.0 Kpc \citep[][]{2005MNRAS.357..304H} we get  $R$ = 31.0 $\pm$ 1.1 \rsun. We notice that the longest wavelength 
flux, corresponding to a magnitude of a filter centered in 22 $\mu$m obtained by the Wide-field Infrared Survey Explorer (WISE) \citep{2010AJ....140.1868W}, is probably an artifact, since no star appears in the coordinates of the object in the corresponding WISE image. 

We calculate a visual extinction of $A_V$ = 3.1 $\times$ $E(B-V)$ = 0.155 $\pm$ 0.012 mag. The absolute magnitude is $M_V$ = $V$ - $A_V$ + 5 - 5 log d =  -5.23 $\pm$ 0.02 mag, considering the average $V$ = 13.84 (Table 1). 
Using a bolometric correction of $BC$ = -2.21 $\pm$ 0.15 for $T_{eff}$ = 23 $\pm$ 1.5 kK  \citep[][]{1996ApJ...469..355F},
we get a bolometric magnitude of $M_{bol}$ = -7.44 $\pm$ 0.15 and a bolometric  luminosity of $L_{bol}$ = 10$^{4.87 \pm 0.06}$ $L_{\odot}$.

The mass of  AzV322 is 16 $\pm$ 1 \msun, as determined from its position in the HR diagram and the evolutionary tracks by \citet{2001A&A...373..555M} for rotating stars with a low metallicity of $Z$ = 0.002, proper for the Small Magellanic Cloud.

In order to  estimate the rotational period of the star  $P_{rot}$ (in days) we  the basic kinematics formula:\\

\begin{equation}
 \frac{r}{R_{\odot}} =  \frac{P_{rot} v}{50.633},
\end{equation}

\noindent
where $r$ is the distance to the centre of rotation and $v$ the linear velocity of material at the equator of the star measured in km s$^{-1}$.
 Using $r$ = 31.0  \rsun, $i$ $\geq$ 30\dg and the mean projected rotational velocities for B0-B2 stars of luminosity class I in the Galaxy, viz.\, 69 $\pm$ 7 km s$^{-1}$ \citep{2002ApJ...573..359A} we get  $P_{rot}$  $\geq$ 11 days.
We see that the short cycle is too small to be the rotational period of the star. Also half of the rotational period (due for instance to two stellar spots) is hard to accept  due to the symmetric light curve and its stability through the years. The rotational hypothesis is also weak due to the presence of additional frequencies in the light curve.





\section{Discussion}


\subsection{The disk}

The double peak emission and the Balmer progression with smaller peak separations at the lower members of the Balmer series suggest
that the shape of the line emitting circumstellar envelope is disk-like  \citep{2013A&ARv..21...69R}. The disk can also be responsible for the infrared excess, as usual for Be star disks, through the scattering of the stellar light by free electrons in the envelope  whose visibility is proportional to the disk projected area and hence better at intermediate latitudes
 \citep{1988A&A...194..167D}. It is surprising that AzV322, being a B0.5 supergiant, does not show the typical emission line pattern of an outflowing radiatively driven wind, i.e. a set of P-Cygni profiles, as usually observed in these systems \citep{2007A&A...463.1093L}. We  notice that the cause for this might be a rapid rotation, which should be able to sustain a circumstellar disk like in Classical Be stars.
 A disk formed by mass transfer in a close binary is other possible hypothesis, but no sign of a {\it close} binary is found, see below.
 
Assuming an optically thin Keplerian disk we can determine the relative velocities of the line forming regions as given in Table\,5, following the relationship \citep{1972ApJ...171..549H}:\\

\begin{equation}
 \frac{\Delta \lambda}{2\, sin\,i} = \sqrt{\frac{GM}{r}},
\end{equation}

\noindent
where $\Delta \lambda$ is the peak separation in velocity units and $r$ the radius of the disk emitting region.
 We find that the H$\alpha$, H$\beta$, H$\gamma$ and  Paschen disks are 6.7, 2.4, 1.5 and 1.2 times larger than the disk region forming He\,I\,5875, respectively. Considering the stellar mass derived in Section 4.3 we determine absolute extensions for disk line emitting regions of 6.4  sin$^{2}$\,i $R_{\star}$ (202 sin$^{2}$\,i  \rsun) for the He\,I\,5875 line forming region and 42.8  sin$^{2}$\,i $R_{\star}$ (1349 sin$^{2}$\,i \rsun) for the H$\alpha$ line forming region. For intermediate latitudes (e.g. $i$ = 60\dg, sin$^{2}$\,i = 0.75),  we obtain a H$\alpha$ disk of radius 32.1 $R_{\star}$, which is larger than typical Be star H$\alpha$ disks, whose radial extension are  3--16  $R_{\star}$ \citep{2013A&ARv..21...69R}. It is possible that this can be explained by the larger budget of UV photons available in the supergiant able to ionize a larger fraction of the circumstellar envelope than in classical Be stars. On the other hand,  the long-term photometric variability shown in Fig.\,4, of  0.1 mag amplitude in 17.5 years in the $I$-band,  is typical for Be stars and  possibly explained by electron scattering in circumstellar envelopes of variable density \citep{1990A&A...230..380D}.
 
\subsection{On the short period and g-mode  pulsations}

As mentioned above, the main photometric period of AzV322 is shorter than the stellar rotational period, hence it cannot be 
described by rotation of spots or some kind of surface activity. Also, as said before, the rotational hypothesis cannot explain the presence of multiple 
photometric frequencies. On the contrary,  
the stability of the periodicities through the years (apart from the variability for $f_1$ timings discussed below) suggests an explanation in terms of stellar pulsations as we show in this section.


AzV322 fits well  the luminosity-$T_{eff}$ plane of periodically variable B-type supergiants (Fig.\,11). These have been interpreted in terms of non-radial gravity-mode oscillations \citep{2006ApJ...650.1111S, 2007A&A...463.1093L, 2015MNRAS.447.2378O} and this can be a possible interpretation for the periodicities found in  AzV322.





The main and secondary photometric periods of AzV322 and their amplitudes fit well the parameters of other B-type supergiants showing periodically variable oscillations 
(Fig.\,12).  This finding strengths even more the interpretation of the photometric periodicity in terms of g-mode pulsations. 

 \begin{figure}
\scalebox{1}[1]{\includegraphics[angle=0, width=8cm]{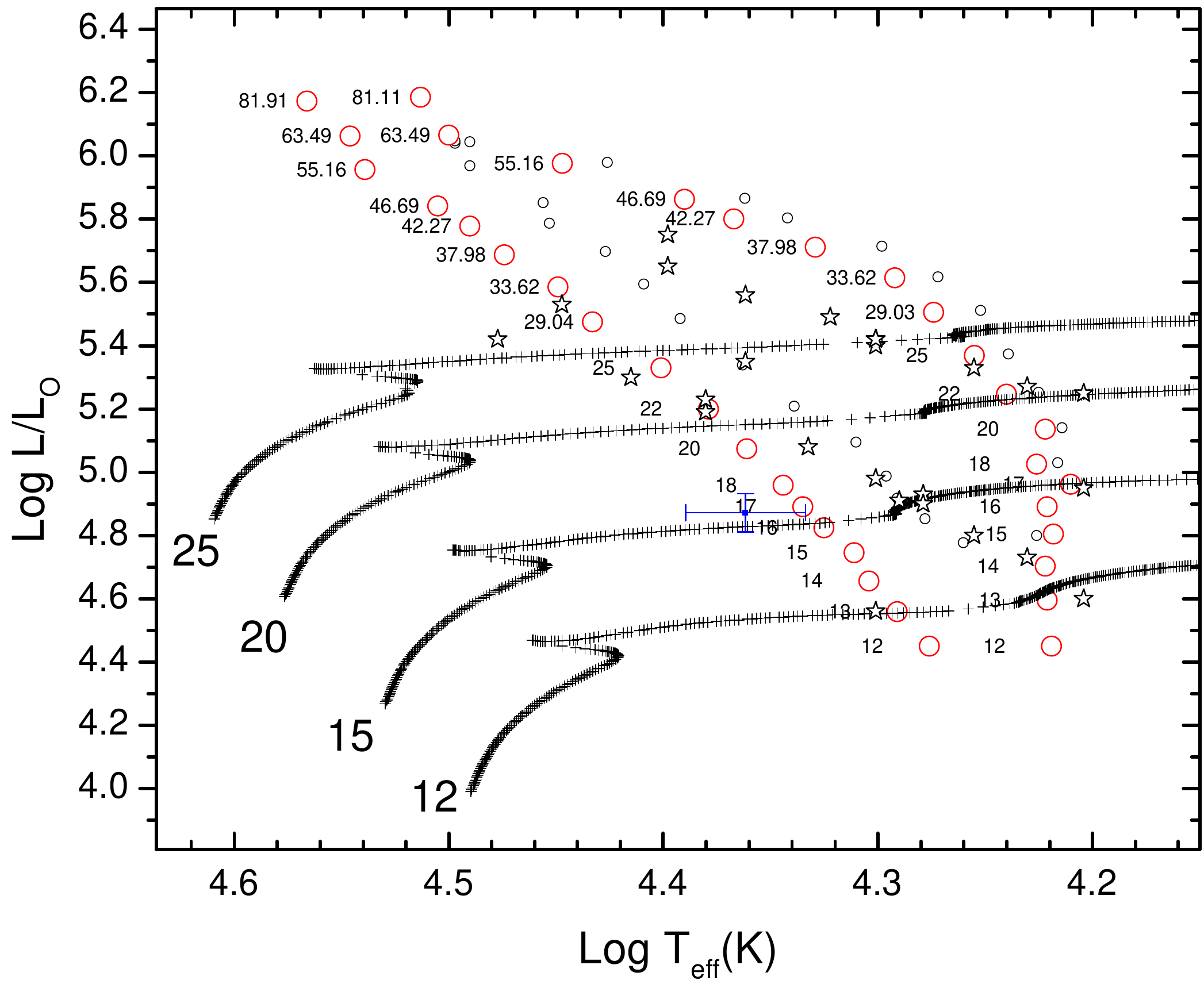}}
\caption{The position of AzV322 in the HR diagram along with the instability boundaries for $l= 1$ (small circles) and $l= 2$ (big circles) gravitationally driven models for metallicity $Z$ = 0.004  of  \citet{2006ApJ...650.1111S}. Some evolutionary tracks from \citet{2001A&A...373..555M} are also shown along with 
the position of periodically variable supergiants (stars) according to  \citet[][]{2007A&A...463.1093L}.Tracks and boundaries are labeled in solar masses.}
  \label{x}
\end{figure}


\begin{figure}
\scalebox{1}[1]{\includegraphics[angle=0, width=8cm]{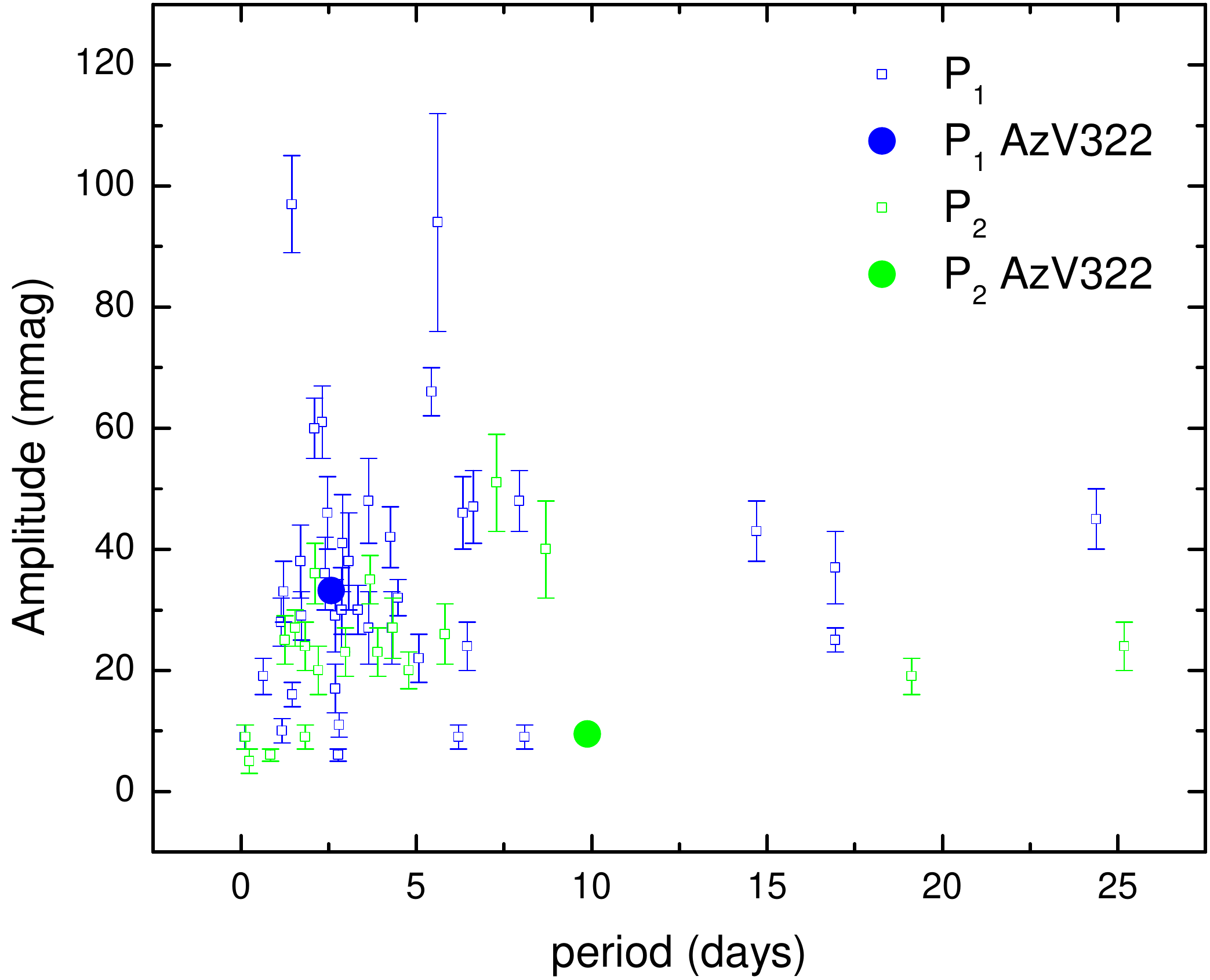}}
\caption{Comparison with the periodically variable B-type supergiants of \citet{2007A&A...463.1093L}. Main and secondary periods and their amplitudes are shown.}
  \label{x}
\end{figure}

We notice that the observed changes in times of maximum observed in the O-C diagram for the frequency $f_1$ might be interpreted in terms of 
a light-travel time effect in a wide-orbit binary. This effect was first suggested for the Algol system by \citet{1888AJ......7..165C} and theoretical tools for its analysis were developed by \citet{1922BAN.....1...93W} and \citet{1952ApJ...116..211I}. It has been noticed that the solution of the inverse problem in which the best solution is determined still remains a problem \citep[][]{2016A&A...589A..94L}. It is beyond the hope of this paper to inquire on the solution 
of this particular phenomenon since a detailed analysis should require data distributed on several decades. It is enough to say here that
if the variability is due to the light-travel time effect, it should indicate a very wide-orbit and long-period binary. The differences in phase of the order of 0.1 cycles for $f_1$ imply time differences of about 0.25 days implying a light travel path of 6.475E9 km i.e.  9308 \rsun, or in terms of the stellar radius 295$R_{\star}$.
This should correspond to the projected semi-major axis of the barycentric orbit of the B-type star, about 9 times the extension derived for the H$\alpha$ emitting disk.

\subsection{Does the long period reflect a binary motion?}

The fact that the centroid of the  H$\alpha$ emission line varies with the long periodicity ($f_2$) suggests that it could represent the period 
of a binary star in an elliptical orbit. The H$\alpha$ is relatively strong and we can easily determine its radial velocity at every epoch, which is harder for the weak and noisy photospheric Helium lines. Actually, helium line RVs show an erratic pattern when phased with $f_2$.

Let's investigate if the  H$\alpha$ radial velocities can be represented by the orbital motion of a binary. We use the genetic algorithm \texttt{PIKAIA}  \citep{1995ApJS..101..309C} to determine
the orbital parameters that best fit the available data. Following the analysis described by \citet{2012MNRAS.427..607M, 2015MNRAS.448.1137M} we minimize $\chi^2$ defined as:\\

\begin{equation}
   \chi^2(P,\tau,\omega,e,K,\gamma) =\frac{1}{N-6}\sum_{j=1}^{n}\left(\frac{V_J-V(t_j,P,\tau,\omega,e,K,\gamma)}{\sigma_j}\right)^2,
\end{equation}

\noindent
where $N$ is the number of observations, $V_j$  and $V$ the observed and calculated velocities in time  $t_j$. The theoretical velocity is:\\

 \begin{equation}\label{eqn:vt}
  V(t)=\gamma + K ((\omega+\theta(t)) + e\cos(\omega)),  
 \end{equation}
 
 \noindent
 where  $\theta$ is the true anomaly obtained by solving the equations:\\
 
\begin{equation}
 \tan\left(\frac{\theta}{2}\right) = \sqrt{\frac{1+e}{1-e}}\tan\left(\frac{E}{2}\right),
\end{equation}

\begin{equation}
 E - e \sin(E)=  \frac{2\pi}{P}(t-\tau),
\end{equation}

\noindent
where $E$ is the eccentric anomaly. We fixed the period, constrained the eccentricity between 0 and 1, $\omega$ between $0$  and $2\pi$,  
 $\tau$ between the minimum HJD and this value plus the period, $K$ between $0$ and $(V_{max} - V_{min})$ and $\gamma$ between $V_{min}$  and $V_{max}$.
 

In order to estimate the errors for the results obtained from \texttt{PIKAIA}, we proceeded to calculate the confidence intervals for the region corresponding to 68.26\% of the sample (1$\sigma$) through a routine created in \texttt{PYTHON} for this purpose. The results are given in Table \ref{tab:pikresult}.
Our results reveal a very eccentric orbit ($e$ = 0.655) and the fit produces a good match to the  available data (residuals $\approx$ 1 km s$^{-1}$, Fig.\,13). From the results above and using the formula:

\begin{equation}
 a \sin i = (1.9758 \times 10^{-2}) (1 - e^2)^{1/2} K P\,  \rm{R_{\odot}}, 
\end{equation}

\noindent
where the period $P$ is in days and the half-amplitude of radial velocity $K$ in kilometers per second (e.g. Hilditch 2001) we get $a_1$sin\,i = 1.64 \rsun. This is a very short distance; even considering a conservative value for the inclination angle of 30\fdg0, we get $a$ = 3.29 \rsun.
Considering that the B star is a supergiant with radius $\sim$ 30 \rsun, it is impossible to fit the orbit. We conclude that the radial velocity variability does not 
reflect a binary orbit.

\begin{figure}
\scalebox{1}[1]{\includegraphics[angle=0, width=8cm]{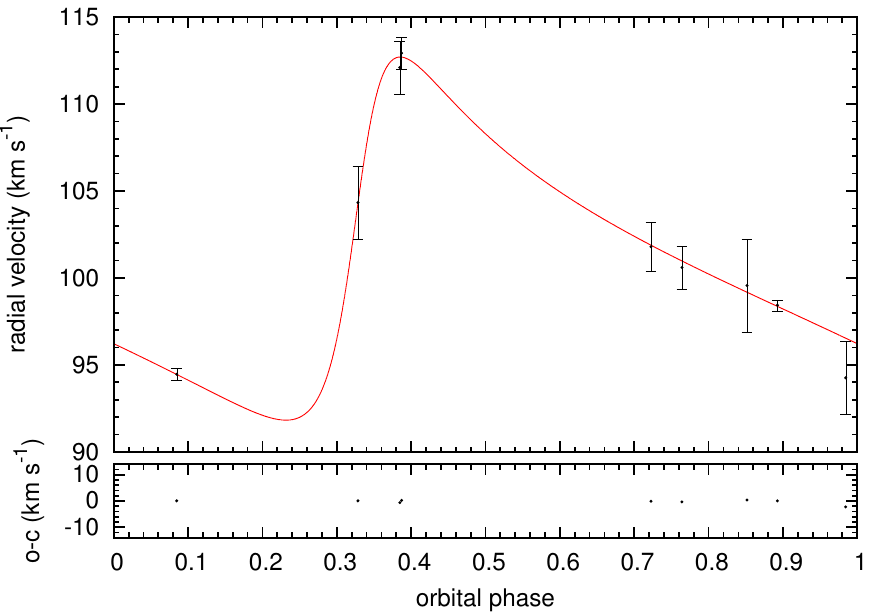}}
\caption{Averaged radial velocities per phase and the best orbital fit.}
  \label{x}
\end{figure}


\begin{table}
    \centering
    \caption{Orbital elements for the donor of AzV322 obtained by minimization of the $\chi^2$ parameter given by Eq.\,4. The value $\tau^*=\tau-2450000$ and the limits of the confidence intervals within one standard deviation ($1\sigma$) are given.}
    \label{tab:pikresult}
    \begin{tabular}{cccc} 
        \hline
        Parameter&Best value&Lower limit&Upper limit\\
        \hline
        $P_o$ (d)   & $9.88370$ (fixed)&  &  \\
        $\tau^{*}$    &4010.71631&4010.71802&4020.55615\\
        $\omega$ (rad)&4.96994&4.91379&5.03079\\
        $e$    &0.59115&0.54371&0.63985\\
        $K_2$ (km s$^{-1}$)&10.44027&10.07962&11.02156\\
        $\gamma$ (km s$^{-1}$)&100.69585&100.42592&100.98771\\
        \hline
        $\chi^2$&&0.51479&\\
        \hline
    \end{tabular}
\end{table}


Since the binary hypothesis has been rejected, other phenomenon produces the modulation of the emission lines radial velocities.
The number of data points is low and more epochs are desirable to define the shape of the radial velocity (RV) curve since it is still possible that the observed variability is product of stochastic phenomena and the low number statistics.

Looking at the Fig.\,12, it is clear that the period of 9\fd8837  ($f_2$) fits the period-amplitude distribution of periodically variable B-type supergiants, so it could be a second mode of gravitationally driven pulsation.

 The origin of the circumstellar envelope at present is unknown, it could be due to: (i)  an unseen companion transferring mass onto the B-type star due to Roche lobe overflow or (ii)
a mechanism related to  g-mode pulsation, in a similar way that seems to be happening in Be stars. At present, the data is not enough to discriminate between these competent scenarios. In particular, more epochs and higher S/N spectra are needed to check the presence of a companion.



 


 \section{Conclusions}

Based on the study of 17.5 years of $I$ and $V$-band OGLE photometry, high-resolution optical spectra covering 8 epochs, and
published photometric magnitudes from the ultraviolet to the infrared region we find that:

\begin{itemize}

\item AzV322 is an early B-type periodically variable supergiant surrounded by a circumstellar disk.

\item  Our study confirms the main photometric periodicity of 2\fd587000 found by \cite{2014A&A...562A.125K}.

\item The star shows  additionally three significant photometric periods:  9\fd8837,  2\fd0503 and 1\fd1438. Both the main photometric  periodicity as well as these three additional frequencies can be attributed to g-mode pulsations. This conjecture is supported by the position of the star in the HR diagram very close to the recently discovered Slowly Pulsating B-type Supergiants by  \citet{2006ApJ...650.1111S} and also for the location of the star in the period-amplitude diagram.

\item The star is characterized by double emission lines and infrared excess consistent with the presence of a circumstellar disk. If optically thin, this disk has extensions 7.6  sin$^{2}$\,i $R_{\star}$  for the He\,I\,5875 line forming region and 50.8  sin$^{2}$\,i $R_{\star}$ for the H$\alpha$ forming region. 

\item The data are compatible with the  following stellar parameters: $T_{eff}$ = 23\,000 $\pm$ 1500 K, log\, g = 3.0 $\pm$ 0.5, $M$ = 16 $\pm$ 1 \msun, $R$ = 31.0 $\pm$ 1 \rsun,  and  $L_{bol}$ = 10$^{4.87 \pm 0.06}$ $L_{\odot}$. 


\item The origin of the disk is unknown, it could be related to the presence of g-mode non-radial pulsations, as happens in Be stars, or alternatively, it could be the result of mass transfer from a Roche lobe filling undetected secondary star.

\item Times of maximum for the $f_1$ frequency vary quasi-cyclically in a time scale of 20 years. This might be interpreted as a light-travel time effect in a wide orbit binary whose secondary stellar component has not been detected.

\end{itemize}

\section{Acknowledgments}

 Thanks to the anonymous referee for providing useful comments that improved the first version of this manuscript.
We thanks Dr. Hideyuki Saio for providing the theoretical data shown in Fig.\,11. This publication makes use of  VOSA, developed under the Spanish Virtual Observatory project supported from the Spanish MICINN through grant AyA2008-02156.
This research has made use of the SIMBAD database, operated at CDS, Strasbourg, France.
R.E.M. acknowledges support by VRID-Enlace 216.016.002-1.0 and the BASAL Centro de Astrof{\'{i}}sica y Tecnolog{\'{i}}as Afines (CATA) PFB--06/2007. 
HEG gratefully acknowledges the financial support by CNPq No. PDJ-152237/2016-0.
The OGLE project has received funding from the Polish National Science
Centre grant MAESTRO no. 2014/14/A/ST9/00121. Thanks to Maja Vuckovic, Monika  Jurkovic and Angelica Jara for  useful conversations during the first stages of the preparation of this manuscript.

\bsp 
\label{lastpage}

\newpage

\end{document}